# Revisiting the work "Brownian motion with time-dependent friction and single-particle dynamics in liquids" by Lad, Patel, and Pratap [Phys. Rev. E 105, 064107 (2022)]

Vladimír Lisý and Jana Tóthová

Department of Physics, Faculty of Electrical Engineering and Informatics, Technical University of Košice, Park Komenského 2, 042 00 Košice, Slovakia

**Abstract**

Recently, Lad, Patel, and Pratap (LP&P) [Phys. Rev. E 105, 064107 (2022)] revisited a microscopic theory of molecular motion in liquids, proposed by Glass and Rice [Phys. Rev. 176, 239 (1968)]. Coming from this theory, LP&P derived a new equation of motion for the velocity autocorrelation function (VAF) and argued that the friction coefficient of particles in liquids should exponentially depend on time. The numerical solution of this equation was fitted to the results of molecular dynamics simulations on different liquids. In our Comment [Phys. Rev. E 108, 036107 (2023)], we showed that this solution, obtained under the condition of zero derivative of the VAF at time $t = 0$, is physically incorrect. This was evidenced by our exact analytical solution for the VAF, not found by LP&P, and numerically, by using the same method as in the commented work. In the Reply [Phys. Rev. E 108, 036108 (2023)], Lad, Patel, Pratap, and Pandya claimed that our solution does not satisfy all the necessary boundary conditions and is thus not appropriate for the description of atomic dynamics in liquids. Until and unless proven otherwise they do not find any reason for the reconsideration of their theory. Here we give a rebuttal to this Reply and, returning to the original work by LP&P, show that the presented there equation for the VAF is wrong. Due to errors in its derivation, it is, among other inconsistencies, incompatible precisely with the boundary conditions for the VAF which lie in the basis of their theory.

## I. Comment on "Brownian motion with time-dependent friction and single-particle dynamics in liquids" [Phys. Rev. E 108, 036107 (2023)]

In this section, we reproduce our comment on the article by Lad, Patel, and Pratap (LP&P) [1]. It contains an introduction to the problematics and allows us in Section II to directly refer to the points discussed in Reply to "Comment on 'Brownian motion with time-dependent friction and single-particle dynamics in liquids' " by Lad, Patel, Pratap, and Pandya (LPP&P) [Phys. Rev. E 108, 036108 (2023)].

In Ref. [1], the authors revisit the theory by Glass and Rice (G&R) [2] of the molecular motion in classical monoatomic liquids. G&R introduced a formalism that was utilized to calculate the velocity autocorrelation function (VAF) of a particle in a liquid. They assumed that the fluctuations arising from the motion of the molecules in the long-range soft part of the intermolecular field are sufficiently rapid to result in an irregular Brownian motion. In addition, G&R represented the effects of the strong short-ranged repulsive core collisions by a time-dependent average force field. The resulting equation of motion corresponds to the classical Langevin equation with the frictional force proportional to the instantaneous velocity $\vec{v}(t)$ of the particle with a constant friction coefficient $\beta$ and an additional systematic force. The obtained equation is transformed into the equation for the normalized VAF, $\psi(t) = \langle \vec{v}(0)\vec{v}(t)\rangle / \langle v^2(0)\rangle$,

$$\frac{d^2\psi}{dt^2} + (\alpha + \beta)\frac{d\psi}{dt} + \left(\omega_0^2 e^{-\alpha t} + \alpha\beta\right)\psi = 0, \qquad (1)$$

where $\omega_0$ is a liquid-characteristic frequency associated with the harmonic potential well and $\alpha$ is a molecular relaxation rate. The VAF is subjected to the conditions $\lim_{t\to 0}\psi(t) = 1$, $\lim_{t\to 0} d\psi(t)/dt = 0$, and $\lim_{t\to 0} d^2\psi(t)/dt^2 = -\omega_0^2$. Equation (1) is then solved in [2] assuming $\alpha = \beta$. In Ref [1], it is shown that Eq. (1) yields $\lim_{t\to 0} d^2\psi(t)/dt^2 = -\left(\omega_0^2 + \alpha\beta\right)$ and thus does not satisfy the last of the above limits. It is proposed to resolve this inconsistency (for $\beta \neq 0$ at $t = 0$ and nonzero $\alpha$) by considering the time-dependent friction coefficient $\beta$. With this assumption and employing the conditions at $t \to 0$, the authors derive, instead of (1), the equation (Eq. (21) in Ref. [1])

$$\frac{d^2\psi}{dt^2} + \left(\alpha + \beta_0 e^{-\alpha t}\right)\frac{d\psi}{dt} + \omega_0^2 e^{-\alpha t}\psi = 0, \qquad (2)$$

where $\beta_0$ is the initial value of the time-dependent friction $\beta(t) = \beta_0 e^{-\alpha t}$. Equation (2) is the main feature of the work [1]. Next, the authors use $\alpha$, $\beta_0$, and $\omega_0$ as fitting parameters, solve Eq. (21) numerically and determine these quantities from the best agreement with the molecular-dynamics results for the VAF for various systems at different densities and temperatures. According to Ref. [1], the equation of motion (2) gives an excellent account in a broad range of liquid densities. A better description of the VAF in low-density fluids was demonstrated for $\alpha > 0$, whereas $\alpha < 0$ is inevitable to obtain consistent results for high-density liquids. However, an elaborate quantitative analysis and physical interpretation, especially for the case $\alpha < 0$, is constrained due to the nonavailability of a tangible analytical solution of Eq. (2) [1]. It was concluded that "This is the primary issue that yet remains to be addressed to acquire an in-depth understanding of the time dependence of the dynamical friction and its implications on the dynamical correlations at short times in liquids."

Here we show that Eq. (2) can be solved exactly in terms of the well-known and widely studied special functions. The solution can be obtained in the following way. After changing the variable $t$ to $x = (\beta_0/\alpha)\exp(-\alpha t)$, Eq. (2) becomes

$$x\frac{d^2\psi}{dx^2} - x\frac{d\psi}{dx} + \frac{\omega_0^2}{\alpha\beta_0}\psi = 0. \qquad (3)$$

This is a special case of Kummer's equation or the confluent hypergeometric equation, see, e.g., [3] (Chap. 13) or [4] (Chap. VI)

$$x\frac{d^2\psi}{dx^2} + (\gamma - x)\frac{d\psi}{dx} - a\psi = 0, \tag{4}$$

where $a = -\omega_0^2/\alpha\beta_0$. Two linearly independent solutions to this equation are confluent hypergeometric function $\Phi(a,\gamma;x) = {}_1F_1(a;\gamma;x)$ and $x^{1-\gamma}{}_1F_1(a-\gamma+1,2-\gamma;x)$ [5] (Chap. 8-9, 9.2). In the case $\gamma = 0$ the series representing the former solution diverges and one can choose the other independent solution $e^{x/2}W_{a,-1/2}(-x)$ [4] (Chap. VI, 6.9). Whittaker's function $W$ can be expressed through the more general Meijer $G$ function, $G_{1,2}^{2,0}\left(-x \Big|_{0,\,1}^{1+\omega_0^2/\alpha\beta_0}\right) = e^{x/2}W_{-\omega_0^2/\alpha\beta_0,-1/2}(-x)$ [4] (Chap. V, 5.6). The general solution to Eq. (3) thus can be written as

$$\psi(x) = c_1 x\, {}_1F_1\left(1-\omega_0^2/\alpha\beta_0;2;x\right) + c_2 G_{1,2}^{2,0}\left(-x \Big|_{0,\,1}^{1+\omega_0^2/\alpha\beta_0}\right). \tag{5}$$

The same form of the solution of Eq. (3) is obtained by the Wolfram Mathematica differential equation solver DSolve [6]. The coefficients $c_1$ and $c_2$ can be determined from the used in Ref. [1] initial conditions $\psi(t=0)=1$ and $d\psi/dt = 0$ at $t = 0$ which, for $\psi(x)$, are

$$\psi\left(x = \frac{\beta_0}{\alpha}\right) = 1, \qquad \left.\frac{d\psi(x)}{dx}\right|_{x=\beta_0/\alpha} = 0. \tag{6}$$

However, the VAF must satisfy the physical condition $\psi(t) \to 0$ at $t \to \infty$ corresponding to the loss of correlation with the initial value $\psi(0)$ with the increase of time, which is inconsistent with the solution obeying the conditions (6). So, for $\alpha > 0$ one has at $t \to \infty$ ($x \to 0$) $\lim_{x\to 0} x\,{}_1F_1\left(1-\omega_0^2/\alpha\beta_0;2;x\right) = 0$ but $\lim_{x\to 0} G_{1,2}^{2,0}\left(-x \Big|_{0,\,1}^{1+\omega_0^2/\alpha\beta_0}\right) = 1/\Gamma\left(1+\omega_0^2/\alpha\beta_0\right)$, where $\Gamma(z)$ is the Gamma function [6]. The solution (5) thus can be physically correct only when $c_2 = 0$. For $\alpha < 0$ at $t \to \infty$ ($x \to -\infty$) $\psi(x)$ from Eq. (5) converges to 0. To have the solution applicable for both $\alpha > 0$ and $\alpha < 0$, we put $c_2 = 0$ and use the necessary condition $\psi(x = \beta_0/\alpha) = 1$ that follows from the definition of the normalized VAF. The Meijer $G$ function can be excluded from the solution (5) also because at $\alpha > 0$ it is a complex function and at $\alpha < 0$ because of an irregular behavior making it unusable to describe the systems of interest (see the note on numerical calculations below). Then the solution for $\psi(x)$ has the form

$$\psi(x) = x\frac{\alpha}{\beta_0} \frac{{}_1F_1\left(1-\omega_0^2/\alpha\beta_0;2;x\right)}{{}_1F_1\left(1-\omega_0^2/\alpha\beta_0;2;\beta_0/\alpha\right)}. \tag{7}$$

The behavior of $\psi(x)$ does not agree with the results obtained in Ref. [1]. This is illustrated by the plots of this function in Figs. 1 and 2 using the parameters that are given in Ref. [1] as obtained from the fit of the numerical solution of Eq. (2) to the molecular dynamics simulation for Lennard-Jones fluids at two different densities. At the same time, numerical solutions of Eq. (2) subjected to the conditions $\psi(t)=1$ and $d\psi(t)/dt=0$ at $t=0$ are shown. These solutions are obtained in the same way as in Ref. [1], i.e., by using NDSolve [6]. It is seen that the analytical solution (7) significantly differs from the numerical solutions of Eq. (2). The calculations are presented in Fig. 1 (for $\alpha > 0$) and Fig. 2 ($\alpha < 0$) for the same time range as in [1], up to 3 ps and 1.5 ps, respectively. The slope of $\psi(t)$ is much slower than in Ref. [1] and it converges to a nonzero constant for $\alpha > 0$ ($\approx 0.95$ in this case). The value of $\psi(t)$ at $t = 3$ ps is slightly larger than 0.95, while in [1] it is about 0.25. The calculations in [1] thus should be corrected. However, the most important merit of this figure is the demonstration that the numerical solution does not satisfy the condition $\psi(t) \to 0$ at $t \to \infty$. The numerical solution for the VAF (with the condition $d\psi(t)/dt = 0$ at $t = 0$) shown in Fig. 2 also differs from that presented in Ref. [1] where the minimum of the VAF curve is reached at a larger time and its absolute value is slightly smaller. In this case the limit of $\psi(t)$ at $t \to \infty$ is correct. Note that at $\alpha < 0$ the Meijer $G$ function changes from approximately 11 at $t = 0$ to $-1.6 \cdot 10^{29}$ at $t = 1.5$ ps. It is a rapidly oscillating function with an increasing amplitude that approaches the value $\sim 10^{386}$ at $t$ around 3.4 ps, then it abrupts to $\sim 3 \cdot 10^2$ and slowly goes to 0 as $t \to \infty$ [6]. Such irregular behavior and giant oscillations support the exclusion of this function from the general solution (5).

Figures 1 and 2 evidence that the calculations presented in Ref. [1] possess unreliable parameters $\alpha$, $\beta_0$, and $\omega_0$. We conclude that the equation for the VAF of a Brownian particle in liquids obtained in Ref. [1] assuming the time dependence of its friction coefficient gives incorrect results if solved under the condition $d\psi(t)/dt = 0$ at $t = 0$. This suggests that the otherwise interesting theory [1] with possible important consequences for the physics of fluids and the Brownian motion should be reconsidered.

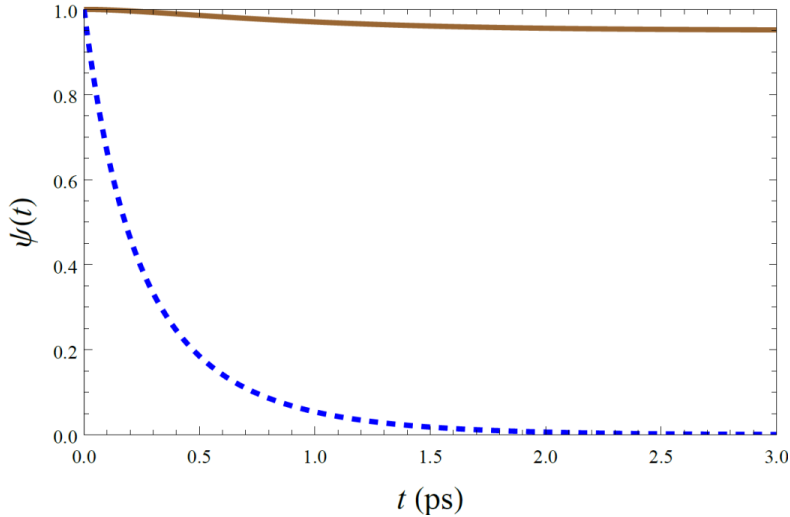

FIG. 1. VAF for Lennard-Jones system calculated numerically from Eq. (2) (full line) subjected to conditions $\psi(0)=1$, $(d\psi/dt)_{t=0}=0$, and from the analytical solution (7) with the condition $\psi(0)=1$ and $\psi(t \to \infty) \to 0$ (dashed line). The parameters are from Ref. [1]: $\alpha = 2.03 \cdot 10^{12}$ s$^{-1}$, $\beta_0 = 3.69 \cdot 10^{12}$ s$^{-1}$, and $\omega_0 = 0.55 \cdot 10^{12}$ s$^{-1}$.

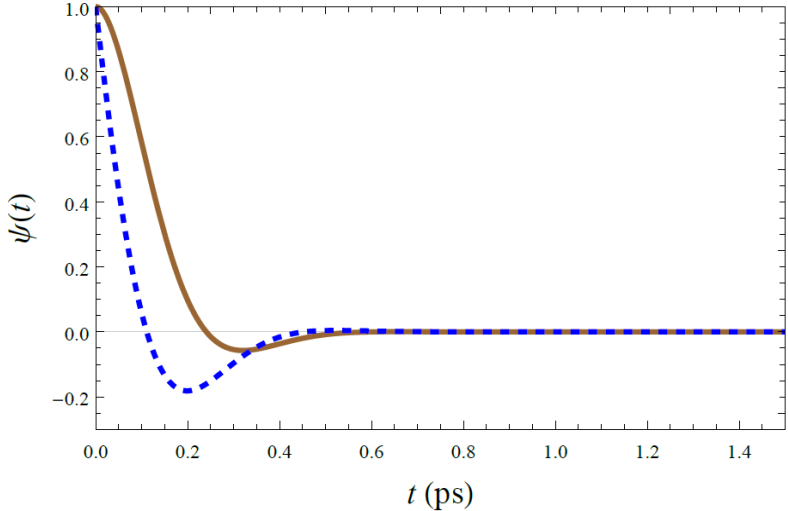

FIG. 2. The same as in Fig. 1 for parameters [1] $\alpha = -1.17 \cdot 10^{12}$ s$^{-1}$, $\beta_0 = 14.79 \cdot 10^{12}$ s$^{-1}$, and $\omega_0 = 11.71 \cdot 10^{12}$ s$^{-1}$.

## II. Rebuttal of Reply to "Comment on 'Brownian motion with time-dependent friction and single-particle dynamics in liquids' " [Phys. Rev. E **108**, 036108 (2023)]

The above-reproduced comment [7] on the work [1] was harshly criticized by LPP&P in the Reply [8], which however contains nothing relevant to challenge our results. This is a case where silence is golden, as the authors have only confirmed the validity of our work. In addition, thanks to the Reply, we were prompted to take a closer look at the work [1] and found it completely incorrect. In this section, we briefly consider the Reply and show its irrelevance. Then, in Section III, we will show the erroneousness of the theory by LP&P.

In the first part of the Reply, LPP&P repeat the equation of motion for the Brownian particle and the associated boundary conditions derived by G&R [2], describe the motivation to look for possible shortcomings in the original G&R theory and to address it for providing a better description of the molecular motion in liquids. The subsequent investigations were reported in Ref. [1] on which we have commented [7]. The major result of their work is the derivation of the equation of motion (2), the boundary conditions for which remain the same as in the work by G&R. LPP&P claim that they attempted to obtain an analytical solution of Eq. (2) but it was observed that the possible general solution invariably involves special mathematical functions which make the determination of all the involved arbitrary constants subject to the boundary conditions (here given after Eq. (1)) physically intractable. Simply put, the authors could not find an analytical solution and opted for a numerical solution of Eq. (2) subject to the given boundary conditions. As distinct from LPP&P, in Ref. [7] we have found an exact analytical solution of Eq. (2). LPP&P repeat our obtaining of the general solution (5) (they have a misprint in it not important for what follows) and write that the knowledge of the coefficients $c_1$ and $c_2$ is necessary to obtain a complete solution $\psi(x)$. We also knew this elementary thing and LPP&P described how these coefficients were determined by us, which led to the solution (7). Then, LPP&P highlight three main points of our Comment:

(1) The parameters $\alpha$, $\beta_0$, and $\omega_0$ given in Ref. [1] are unreliable.

(2) The equation of motion (2) gives incorrect results if solved under the condition $d\psi/dt = 0$ at $t = 0$.

(3) The interesting theory [1] should be reconsidered.

The refutation of point (1) LPP&P begin by admitting that their parameters $\beta_0$ and $\omega_0$ for the LJ system that we used in our illustrative calculations (see FIG. 1) are wrong but all other reported values are correct. They claim that "the fact that the results obtained using the numerical solution of Eq. (2) are in close agreement with the classical molecular dynamics (MD) results for a variety of liquids with a wide range of density is a testimony to the reliability of the parameters." By all means the successful fit of the numerical solution with three arbitrary parameters to the MD results cannot serve as a testimony of the reliability of these parameters. Many other solutions not less substantiated can be used to get an agreement with the MD. For us, it was senseless to check all 21 results of VAFs reported by LP&P for a variety of liquids since it was clear that the solutions presented by LP&P cannot be correct if they do not satisfy the necessary condition for the VAF at $t \to \infty$ (the condition ignored in Ref. [1] and not properly discussed even in the Reply [8]).

As to the remarks (2) and (3), LPP&P claim that they are unjustified because the derivation of the analytical solution, Eq. (7), is flawed from the viewpoint of physics involved in Eq. (2). In the next section we will prove that Eq. (2) is wrong. Here we will discuss the pointless explanation by LPP&P of why the general solution (5) cannot possess the solution of the equation of motion of a Brownian particle in a liquid. LPP&P first agree that there is no connection about the solution (5) and once more mention that they were also aware of the analytical solution of Eq. (2) in terms of special functions. However, a complete and correct solution warrants rigorous determination of the arbitrary constants $c_1$ and $c_2$, and for this it is essential that any special function that is a part of the general solution should be a regular and continuous function in the given range of time. This is not true. Yes, the final solution describing a physical system must, after the determination of the arbitrary constants, obey these requirements and if the general solution contains a special function with "bad" properties, this part must be excluded from the solution. This was exactly what we did in Ref. [7]: the Meijer $G$ function was excluded from the general solution (5). A large part of the Reply is therefore totally useless as it is devoted to the properties of the Meijer $G$ function, which are mentioned also in our Comment [7]. In addition, the physics of the problem requires that $\psi(t) \to 0$ at $t \to \infty$. Consequently, see the preceding section, the Meijer $G$ function cannot be a part of the solution, irrespective of other conditions on the VAF. The authors forcefully require that the boundary conditions for $\psi(t)$ at $t = 0$ be met but it is impossible after we put $c_2 = 0$. These conditions are no longer needed and the claim about the inappropriateness of our solution, which does not meet these conditions, is meaningless. The effort of LPP&P to show that the first and second derivatives of our solution (7) do not satisfy these conditions is therefore useless.

We mention the even more bizarre remark by LPP&P that we overlooked the checking of the validity of equation (7) by comparing it with the MD results. Properly, the authors should ask themselves why they did not use our solution (7) to determine the three parameters contained therein by using a fit to the MD results. This concerns also Fig. 2 in Ref. [8] where $\psi(t)$ is presented as calculated with the corrected by LPP&P parameters from our solution, from the numerical solution of Eq. (2), and the MD. According to LPP&P, the new parameters $\beta_0$ and $\omega_0$ are $5.5 \times 10^{12}$ s$^{-1}$ and $36.9 \times 10^{12}$ s$^{-1}$, respectively [8]. With such parameters $\psi(t)$ is a damped oscillating function (Fig. 3) that has nothing in common with the monotonically decaying MD curve. We changed the values to $\beta_0 = 36.9 \times 10^{12}$ s$^{-1}$ and $\omega_0 = 5.5 \times 10^{12}$ s$^{-1}$ and got curves that correspond to those in Fig. 2 (a) of Ref. [8]. This Fig. 4 again indicates that the numerical solution at $\alpha > 0$ does not satisfy the condition $\psi(t) \to 0$ at long times which is ignored by LPP&P. Naturally, the shown there curve obtained from the analytical solution (7) cannot agree with the numerical solution of Eq. (2) obtained without this condition and fitted to the MD simulations.

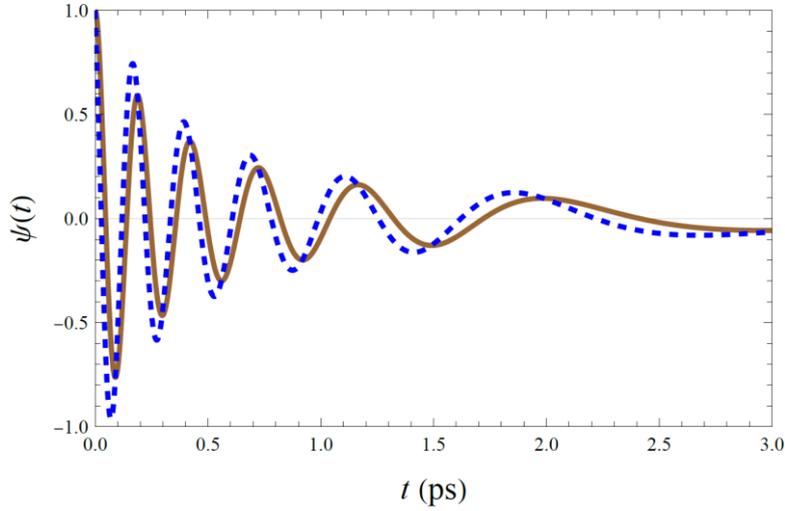

FIG. 3. The same as Fig. 1 with the new parameters from Ref. [8]: $\alpha = 2.03 \cdot 10^{12}$ s$^{-1}$, $\beta_0 = 5.5 \cdot 10^{12}$ s$^{-1}$, and $\omega_0 = 36.9 \cdot 10^{12}$ s$^{-1}$.

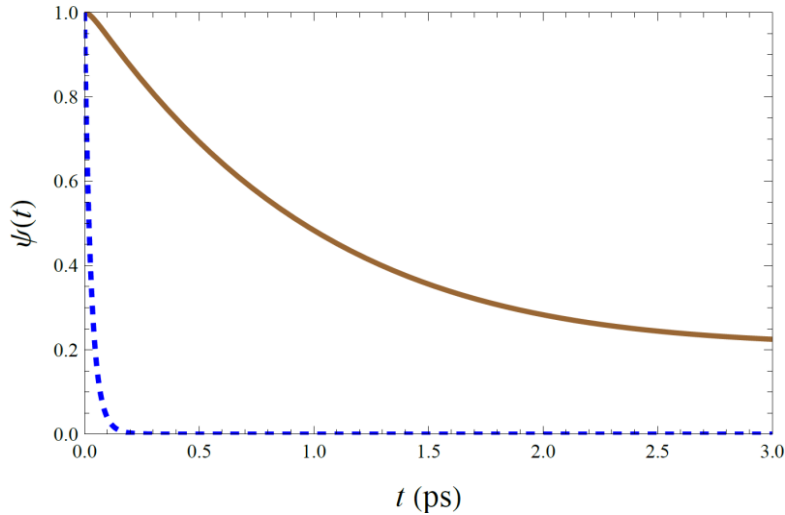

FIG. 4. The same as in Figs. 1 and 3 for parameters $\alpha = 2.03 \cdot 10^{12}$ s$^{-1}$, $\beta_0 = 36.9 \cdot 10^{12}$ s$^{-1}$, and $\omega_0 = 5.5 \cdot 10^{12}$ s$^{-1}$ (we swapped the values of $\beta_0$ and $\omega_0$ against those given in Ref. 8).

To summarize this section, it should be clear that LPP&P do not have adequate arguments to disprove our comment [7]. Moreover, in the next section, we will show that the equations for VAF used in Ref. [1] are inconsistent with the boundary conditions underlying the entire theory of LP&P.

**III. Errors in the derivation of the equation for the VAF by Lad et al. [Phys. Rev. E 105, 064107 (2022)]**

Our comment [7] presented in Section I was not directed against the model of LP&P. The obtained by us solution of the equation for the VAF was supposed to help the further development of their theory, which seemed to be meaningful. However, after a surprisingly unfounded attempt to refute our solution, we returned to the theory presented in the original article [1]. Checking the derivation of the equation for VAF led us to the provable conclusion that Eq. (2) (Eq. (21) in Ref. [1]) obtained as the "main feature" of their work is wrong. It can be easily shown that this equation does not agree with the conditions for the VAF at $t = 0$, on which the whole theory is built.

Following the theory by G&R [2], LP&P consider the equation (Eq. (13) in [1])

$$\frac{d\psi}{dt} = -\beta\psi - \omega_0^2 e^{-\alpha t}\int_0^t dt'\psi(t') . \qquad (8)$$

In Ref. [2], the friction coefficient $\beta$ is a time-independent parameter. In this case, however, Eq. (8) fails for the times $t \to 0$: for nonzero $\beta$ and $\psi(0) = 1$ it gives $\lim_{t\to 0} d\psi(t)/dt = -\beta$) while, according to LP&P, this limit should be zero. Unlike them, G&R understand that a meaningful concept for the friction coefficient is only for times that are long compared to the period of the rapidly fluctuating forces. Therefore, as $t \to 0$, Eq. (8) ceases to be valid.

We can also note that LP&P did not derive Eq. (8) correctly. They came from the equation for the velocity of the Brownian particle [2],

$$m\frac{d\vec{\upsilon}}{dt} = -m\beta\vec{\upsilon} + \vec{F}(\vec{R},t) + \vec{A}(t), \qquad (9)$$

where $\vec{A}(t)$ is a fluctuating force for which G&R explicitly assume that $\langle \vec{A}(0)\vec{A}(t)\rangle = 0$ and $\langle \vec{\upsilon}(0)\vec{A}(t)\rangle = 0$, and $\vec{F}(\vec{R},t) \approx -m\omega_0^2\vec{R}(t)e^{-\alpha t}$ is an ensemble averaged force acting on a particle which was at (0, 0) and has moved to $(\vec{R},t)$ so that $\vec{R}(t) = \int_0^t \vec{\upsilon}(t')dt'$. LP&P multiply Eq. (9) by $\vec{\upsilon}(0)/\upsilon^2(0)$ and average. However, the correct way to obtain Eq. (8) is to multiply Eq. (9) by $\vec{\upsilon}(0)/\langle\upsilon^2(0)\rangle$ and then average the result. Acting as LP&P, we would have to assume that $\langle\vec{\upsilon}(0)\vec{\upsilon}(t)/\upsilon^2(0)\rangle = \langle\vec{\upsilon}(0)\vec{\upsilon}(t)\rangle/\langle\upsilon^2(0)\rangle$ and $\langle\vec{\upsilon}(0)\vec{F}(\vec{R},t)/\upsilon^2(0)\rangle = \langle\vec{\upsilon}(0)\vec{F}(\vec{R},t)\rangle/\langle\upsilon^2(0)\rangle$.

Except for a few typos, up to Eq. (8) (which is incorrect if one insists on the condition $d\psi/dt = 0$ at $t = 0$), the theory by LP&P [1] is the same as in Ref. [2]. Differentiating Eq. (8)

with respect to time, in both papers Eq. (1) is found. G&R obtained the solution of this equation obeying a physically reasonable assumption $\alpha = \beta$ and the conditions at $t = 0$. The VAF of G&R obeys also the condition at $t \to \infty$. Their solution can be however questioned since the equation itself does not satisfy the condition for $d^2\psi/dt^2$ at $t = 0$. As mentioned in Section I after Eq. (1), $\lim_{t\to 0} d^2\psi(t)/dt^2 = -(\omega_0^2 + \alpha\beta)$ instead of the assumed condition $\lim_{t\to 0} d^2\psi(t)/dt^2 = -\omega_0^2$. The former limit is obtained under the condition $\lim_{t\to 0} d\psi(t)/dt = 0$ but, as we saw above, Eq. (8), from which Eq. (1) is obtained, gives a different limit: $\lim_{t\to 0} d\psi(t)/dt = -\beta$. A different limit also applies to $d^2\psi/dt^2$ from Eq. (1): $\lim_{t\to 0} d^2\psi(t)/dt^2 = \beta^2 - \omega_0^2$. LP&P thus found the wrong limits from Eqs. (8) and (1), which do not agree with the initial conditions for the VAF. Only when $\beta = 0$ the initial conditions for the first and second derivatives of the VAF are consistent with Eqs. (8) and (1). This is possible in the consideration by G&R but not by LP&P.

The errors connected with the correspondence of the used equations and the boundary conditions appear also in the "generalization" of the G&R theory by LP&P. They believe that the time independence of $\beta$ is inadequate for a situation where a particle moves in a mean time-dependent force field and the strong, short-ranged repulsive collisions are considered to be dynamically correlated. As an attempt to resolve the discrepancy between Eq. (8) and the boundary conditions, the assumption that the coefficient $\beta$ is time-dependent is made. Although the consideration by G&R of constant $\beta \neq 0$ at longer times and $\beta = 0$ at $t = 0$ can be accepted, let us assume together with LP&P that $\beta(t)$ depends on time. LP&P then differentiate Eq. (8) with respect to time to obtain

$$\frac{d^2\psi}{dt^2} + [\alpha + \beta(t)]\frac{d\psi}{dt} + \left[\omega_0^2 e^{-\alpha t} + \frac{d\beta(t)}{dt} + \alpha\beta(t)\right]\psi = 0 \qquad (10)$$

and to get from this equation an expression for $\beta(t)$. However, let us consider the $t \to 0$ limit of Eq. (8) with $\beta(t)$. One immediately obtains

$$\lim_{t\to 0} \frac{d\psi(t)}{dt} = -\beta(0), \qquad (11)$$

which is again inconsistent with the condition $\lim_{t\to 0} d\psi(t)/dt = 0$ except the case $\beta(0) = 0$. This strictly required by LP&P condition thus contradicts their own theory.

Let us now see how the expression $\beta(t) = \beta_0 e^{-\alpha t}$ was derived by LP&P. Searching for the $t \to 0$ limit of Eq. (10), they (erroneously, due to Eq. (11)) omit the limit of the second term on the left and consider the equation

$$\lim_{t\to 0} \frac{d^2\psi}{dt^2} = -\omega_0^2 \psi - \lim_{t\to 0}\left[\frac{d\beta(t)}{dt} + \alpha\beta(t)\right]\psi. \qquad (12)$$

In order to have $\lim_{t\to 0} d^2\psi(t)/dt^2 = -\omega_0^2$, they determine $\beta(t) = \beta_0 e^{-\alpha t}$ as the solution of the equation $d\beta(t)/dt + \alpha\beta(t) = 0$. It is true that if the latter equation is satisfied, $\lim_{t\to 0}[d\beta(t)/dt + \alpha\beta(t)] = 0$ as well. However, this limit will be equal to zero also in infinitely many other cases, e.g., when $\beta(t) = bt^k$, $k \geq 2$. The expression $\beta(t) = \beta_0 e^{-\alpha t}$ (we used it in our recent paper on the dynamics of a Brownian particle in fluids with time-dependent friction [9, 10]) is only a particular possible solution that leads to the required by LP&P boundary condition for $d^2\psi/dt^2$. At the same time, due to Eq. (11), this condition is not satisfied by the LP&P solution for $\beta(t)$: from Eq. (10) with $\lim_{t\to 0} d\psi(t)/dt = -\beta_0$ one obtains

$$\lim_{t\to 0}\frac{d^2\psi}{dt^2} = -\omega_0^2 + \beta_0(\alpha + \beta_0) \tag{13}$$

instead of $-\omega_0^2$ on the right-hand side.

## IV. Conclusion

In conclusion, we have not only rebutted the Reply [8] by LPP&P to our Comment [7] but also showed that all the theory presented by LP&P in Ref. [1] is wrong.

The Reply contains nothing relevant to contradict our results in the Comment. In fact, the authors have only confirmed its validity. In particular, it concerns the reliability of the parameters $\alpha$, $\beta_0$, and $\omega_0$ given in Ref. [1]. LPP&P admit that their $\beta_0$ and $\omega_0$ for the LJ system that we used in our illustrative calculations (FIG. 1) were wrong (but they claim that all other reported values are correct). We disproved the pointless explanation by LPP&P as to why the general solution (5) cannot possess the solution of the equation of motion of a Brownian particle in a liquid. LPP&P agree that there is no connection about the solution (5), but at the same time, it is wrong since it contains a special function (Meijer $G$) which has properties not suitable for solving the physical problems. However, this argument is irrelevant since in the Comment [7] this function was eliminated from consideration. A big part of the Reply is therefore worthless. The fact that the physics of the problem requires $\psi(t) \to 0$ at $t \to \infty$ is ignored by the authors. LPP&P then have no adequate arguments to challenge our Comment [7].

After a surprisingly unfounded attempt to refute our solution, we returned to the theory presented in the original article [1].

Checking the derivation of the equation for VAF led us to the provable conclusion that Eq. (2) (Eq. (21) in Ref. [1] and Eq. (3) in [8]) named the "main feature" of their work is wrong. It can be easily shown that this equation does not agree with the conditions for the VAF at $t = 0$, on which the whole theory by LP&P is built. The (not derived correctly) basic equation of the theory (Eq. (13) in Ref. [1], Eq. (8) in this article) for nonzero $\beta$ and $\psi(0) = 1$ gives $\lim_{t\to 0} d\psi(t)/dt = -\beta$ while, according to LP&P, this limit should be zero. The main equation of the paper [1], Eq. (21) (Eq. (2) here) also does not agree with the required condition at $t \to 0$.

Finally, attempting to satisfy the condition $\lim_{t \to 0} d^2\psi(t)/dt^2 = -\omega_0^2$, LP&P find $\beta(t) = \beta_0 e^{-\alpha t}$ for the time-dependent friction coefficient. However, this is only one from an infinite number of possible solutions, which is proposed without any substantiation. Furthermore, if Eq. (8) is to hold, the actual initial condition will be different.

To summarize, we have proved that not only the equation of motion (2) gives incorrect results if solved under the condition $d\psi/dt = 0$ at $t = 0$, as we argued in the Comment [7], but the whole theory presented in Ref. [1] is wrong and should be reconsidered.

**Acknowledgment.** This work was supported by the Scientific Grant Agency of the Slovak Republic through Grant VEGA No. 1/0353/22.